\documentclass[twocolumn]{aastex701}

\newcommand\osv{O\,{\sc vii} }
\newcommand\oeit{O\,{\sc viii} }

\usepackage{soul}
\usepackage{amsmath}

\shorttitle{Extended Hot Halo of M\,31}
\shortauthors{Wang et al.}

\begin{document}

\title{An X-ray Absorption-Line Survey of the Circumgalactic Medium of M\,31 with {\it XMM-Newton}}

\author[0000-0002-5879-3360]{Kaile Wang}
\affiliation{Department of Astronomy, Xiamen University, Xiamen, Fujian 361005, China}
\affiliation{Department of Physics, University of Texas at Austin, Austin, Texas 78712, USA}
\email{kailewang@utexas.edu}

\author[0009-0000-3060-1219]{Zheng Zhou}
\altaffiliation{These authors share corresponding authorship.}
\affiliation{Department of Astronomy, Xiamen University, Xiamen, Fujian 361005, China}
\email[show]{zhengz@stu.xmu.edu.cn}

\author[0000-0002-2853-3808]{Taotao Fang}
\altaffiliation{These authors share corresponding authorship.}
\affiliation{Department of Astronomy, Xiamen University, Xiamen, Fujian 361005, China}
\email[show]{fangt@xmu.edu.cn}

\author[0000-0002-6896-1364]{Fabrizio Nicastro}
\affiliation{Istituto Nazionale di Astrofisica (INAF) - Osservatorio Astronomico di Roma Via Frascati 33 00078 \\ Monte Porzio Catone (RM), Italy}
\email{fabrizio.nicastro@inaf.it}

\begin{abstract} 
X-ray absorption line spectroscopy provides a powerful method for probing the extended hot circumgalactic medium (CGM) of individual galaxies. In this study, we present the first survey of \osv and \oeit absorption lines toward the hot halo of M\,31, using $15$ and $18$ background active galactic nuclei, with impact parameters ranging from $R_{\rm imp}\sim300$--$730$~kpc and $\sim190$--$730$~kpc, respectively. We find a marginal excess absorption above the expected Milky Way (MW) foreground toward the innermost sightlines, consistent with an additional hot-CGM contribution associated with M\,31. Comparing sightlines inside and outside $R_{\rm imp}=300$--$350$~kpc, the excess corresponds to a mean equivalent width of $\sim4$--$9$~m\AA\ for the inner \osv sightlines at $R_{\rm imp}\sim310$~kpc and $\sim17$~m\AA\ for the inner \oeit sightlines at $R_{\rm imp}\sim200$~kpc, with a nominal combined significance level of $2.2$--$2.3\,\sigma$. The hot-CGM mass inferred from this excess is highly model dependent, especially on the assumed CGM boundary, and therefore does not yet provide a robust baryon census. Attributing the excess entirely to hot gas confined within approximately the virial radius of M\,31 would require a CGM mass exceeding the nominal ``missing'' baryon budget. The inferred mass decreases to a few $\times10^{11}~M_\odot$ when the assumed CGM boundary is extended to $400-500$~kpc. Deeper observations, both through longer exposures of existing sightlines and the inclusion of additional inner-halo targets, will be essential to robustly constrain the properties of the hot CGM of M\,31.
\end{abstract}


\section{Introduction} 
\label{sec:intro}
The hot circumgalactic medium (CGM) at $T\sim10^6$~K is a leading candidate for hosting a substantial fraction of the ``missing'' galactic baryons (e.g., \citealt{Sommer-Larsen2006, Gupta2012, Fang2013, Nicastro2016, Faerman2017, Das2020, Nicastro2023}). However, its spatial distribution, and therefore the total mass, remain poorly constrained in observation. 

For the Milky Way (MW), reconstructing the radial distribution of the hot CGM is inherently model-dependent, because we observe it from within the Galaxy. Both emission and absorption measurements trace the cumulative contribution of gas along the entire line of sight (LOS), leading to a loss of radial information and a degeneracy between gas density and path length. Consequently, previous studies have inferred a wide range of hot-CGM baryon budgets, from $\sim30\%$ to nearly all of the Galaxy's ``missing'' baryons (e.g., \citealt{Miller2013, Miller2015, Troitsky2017, Li2017, Martynenko2022, Locatelli2024}). External galaxies provide a cleaner geometric perspective, but detecting their hot CGM at large radii (i.e., $\gtrsim0.2~R_{\rm vir}$, where $R_{\rm vir}$ is the virial radius) remains challenging due to the low gas density (see, e.g., \citealt{Ma2015, Anderson2016, Bogdan2017, Li2017ApJS, Singh2018, Comparat2022, Bregman2022}). The radial extent and total mass of this gas component therefore remain highly uncertain.

X-ray absorption studies toward bright active galactic nuclei (AGNs) have recently demonstrated the capability to detect hot CGM out to large radii \citep{Mathur2023, Nicastro2023}, with \osv absorption lines from $L^*$ galaxies being detected at $\sim(0.5-0.6)~R_{\rm vir}$ with high significance levels ($\gtrsim4\,\sigma$). However, a major limitation of this technique is the rarity of X-ray bright AGNs. As a result, these studies rely on a single line of sight to probe the hot CGM of each galaxy, which prevents a detailed investigation of the gas distribution.

The Andromeda galaxy (M\,31, $l=121.174^\circ$, $b=-21.573^\circ$) offers a unique opportunity to overcome this limitation thanks to its large sky coverage. As the closest $L^*$ galaxy, its large angular extent increases the likelihood of identifying multiple bright background AGNs for absorption studies. In the ultraviolet band, absorption from the cool- and warm-phase CGM of M\,31 has been studied along multiple AGN sightlines \citep{Lehner2015, Lehner2020, Lehner2026}. Their findings indicate that the CGM in M\,31 is complex in the gas phase and dynamics at small radii, while becoming more highly ionized in the outer regions. Notably, the most ionized species in their survey, i.e., O\,{\sc vi}, tracing a temperature of $\sim10^{5.5}$~K (e.g., \citealt{Gnat2007}), is detected up to approximately $1.9~R_{\rm vir}$. This result reveals an extended distribution of highly ionized gas surrounding M\,31 and motivates searches for hotter CGM phases at large radii.

Recently, we analyzed the {\it XMM-Newton} spectra of the AGN target PG\,0052+251, whose sightline intersects the hot halo of M\,31 at $\sim 0.7~R_{\rm vir}$ \citep{Zhou2025}. With the absorption features from the $z\sim0$ hot gas and the AGN intrinsic gas explicitly analyzed by plasma models, we identified a $\sim2\,\sigma$ excess in the $z\sim0$ absorption beyond the expected Milky-Way (MW) contribution, likely attributed to the hot CGM of M\,31 along the LOS.

Motivated by this tentative detection, we present the first systematic X-ray absorption survey of the hot CGM in M\,31 with {\it XMM-Newton}, using \osv and \oeit absorption lines. In Section~\ref{sec:data_analysis}, we describe the procedures for sample selection and data analysis. In Section~\ref{sec:results}, we highlight the excess \osv and \oeit absorption toward M\,31 relative to the MW foreground and infer the hot-CGM mass from this excess. In Section~\ref{sec:discussion}, we discuss caveats and systematic uncertainties affecting the interpretation of the observed excess. Our conclusions are summarized in Section~\ref{sec:Summary}. Throughout this paper, we adopt a distance of $D_{\rm M\,31} = 770$~kpc between the MW and M\,31 (\citealp{van2012}; see also \citealt{Li2021_DM31, Pipwala2025} for recent Cepheid measurements, and \citealp{Bhattacharya2025} for a review) and a virial radius of $R_{\rm vir,\,M\,31} = 300$~kpc, corresponding to a virial mass of $M_{\rm vir,\,M31} = 1.51\times10^{12}~M_\odot$ \citep{van2012}. Both quantities remain observationally uncertain, and we address this by exploring different boundary choices for the hot CGM of M\,31 in Section~\ref{sec:results}.

\section{Data Analysis}
\label{sec:data_analysis}

\subsection{Sample Selection}
\label{subsec:sample_select}
Our analysis focuses on the K$\alpha$ absorption lines of \osv (rest-frame wavelength of $21.60$~\AA) and O\,{\sc viii} ($18.97$~\AA), which are optimal tracers of million-degree gas (e.g., \citealp{Gnat2007, Bregman2015JATIS}). We searched the archive of {\it XMM-Newton}/RGS (Reflection Grating Spectrometer, \citealt{den_Herder2001}) for X-ray bright AGNs behind M\,31. Sources with impact parameters of $R_{\rm imp}\leq750$~kpc were considered. This threshold, corresponding to $2.5$ times the virial radius of M\,31, is sufficient to probe the spatial distribution of its hot CGM. We restrict the present survey to {\it XMM-Newton/}RGS data to maintain a homogeneous instrumental response, reduction procedure, and line-spread function. A joint {\it XMM-Newton}/{\it Chandra} analysis is left for future work.

\begin{figure}
\centering
\includegraphics[width=0.98\linewidth]{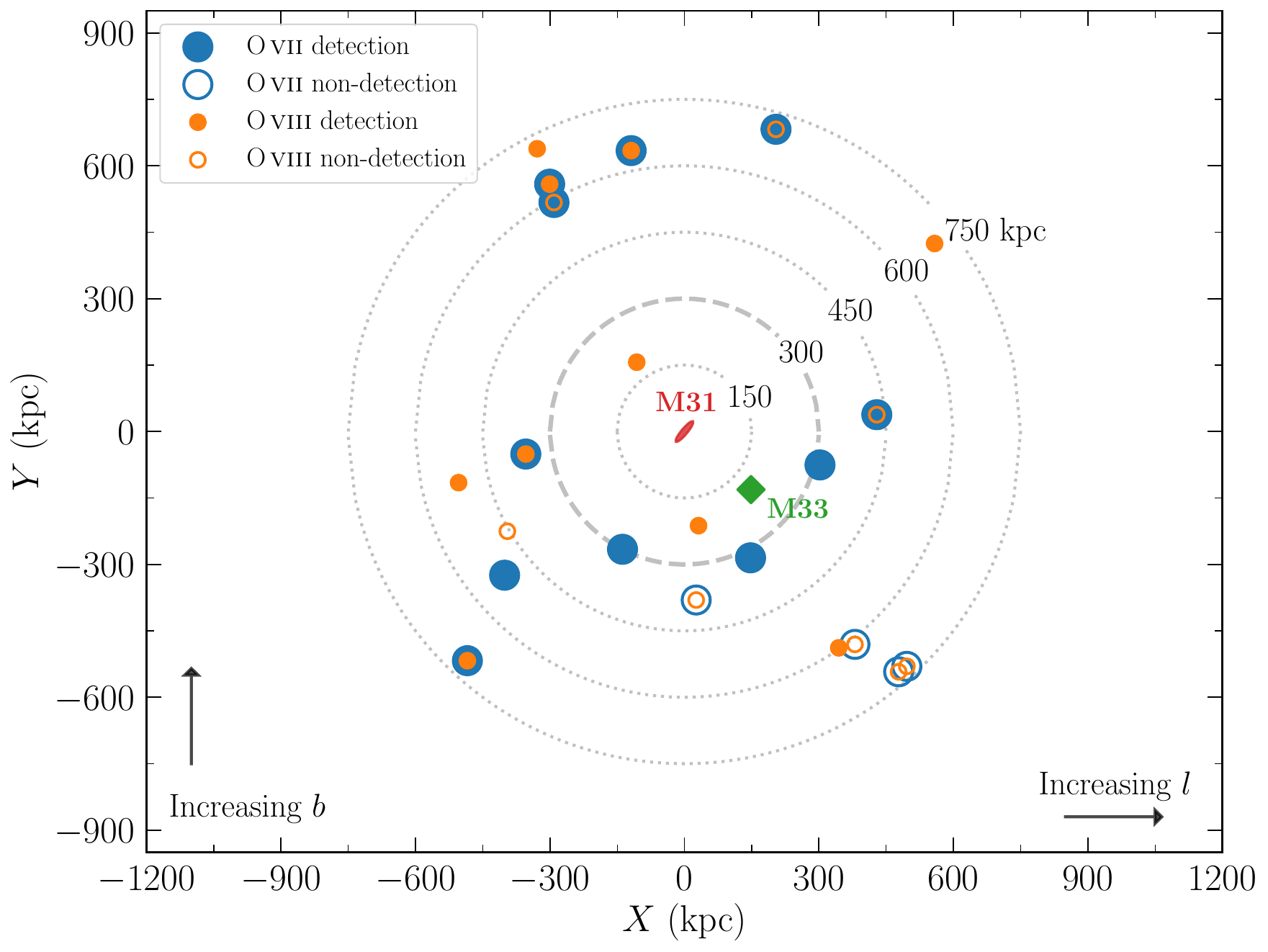} 
\caption{Projected locations of the selected \osv and \oeit samples relative to M\,31. Large blue circles show the \osv sample, while small orange circles show the \oeit sample. Filled and open circles indicate detections and non-detections, respectively. The red ellipse centered at the origin illustrates the projected orientation of the M\,31 disk, and the green diamond marks the projected location of M\,33.
}
\label{fig:targets_location}
\end{figure}

To ensure a reliable analysis of the weak \osv and \oeit lines, we require the counts per resolution element (CPRE\footnote{CPRE is defined as ${\rm CPRE} = f_{\lambda} A_{\rm eff} T \Delta \lambda$. In the equation, $f_{\lambda}$ is the spectral flux, $A_{\rm eff}$ is the effective area of the spectrometer, $T$ is the exposure time, and $\Delta \lambda$ is the defined width for each resolution element. We adopt $\Delta \lambda=25$~m\AA\ in this study.}, e.g., \citealt{Fang2015, Luo2018}) of the selected AGNs to exceed $20$ counts at the spectral continua near $21.60$~\AA\ and $18.97$~\AA\ for selecting the \osv and \oeit samples, respectively. Some targets, such as PG\,0052+251 \citep{Zhou2025}, are known to host AGN intrinsic absorbers or emitters (e.g., \citealt{Tombesi2013, Mao2018, Laha2021}). We have carefully examined the spectrum of each target with results from past studies and retained those LOSs where AGN lines and the $z\sim0$ oxygen lines are not blended. Applying these criteria yielded final samples of $15$ AGNs for \osv and $18$ for \oeit (Appendix~\ref{append:fit_result}). Their impact parameters span approximately $300-730$~kpc and $190-730$~kpc for the \osv and O\,{\sc viii}, respectively. Figure~\ref{fig:targets_location} displays the locations of the selected AGNs, with M\,31 centered at the origin.

\subsection{Data Reduction and Spectral Analysis} 
\label{subsec:spec_analysis}
We extracted the RGS spectra following the standard routine of the {\it XMM-Newton} Science Analysis System (SAS, version~21.0.0), with the updated calibration files. The chain task ``{\it rgsproc}'' was executed to extract the first-order spectra, with cool pixels removed by setting ``keepcool=no''. We generated CCD\,9 light curves and filtered out soft-proton flares using a threshold of ``${\rm Rate}>0.5$~counts\,s$^{-1}$''. The cleaned spectra of each target were then stacked using the SAS tool ``{\it rgscombine}''.

We performed spectral analysis using the X-ray spectral fitting package XSPEC, version 12.14.0h \citep{Arnaud1996}. As broadband features are not the focus of this work, we analyzed the spectral segments of $21-22$~\AA\ for \osv and $18.5-19.5$~\AA\ for O\,{\sc viii}. The spectral range of $21.77-21.84$~\AA\ was excluded due to a known instrumental feature. The continuum in each segment was modeled with a power law absorbed by the Galactic neutral gas ({\it pow*phabs}). The LOS hydrogen column density of the Galactic neutral gas was fixed to the result of the HI4PI survey \citep{HI4PI2016}. 

We modeled the oxygen absorption lines using the Voigt profile developed by \citet{Buote2009}, which involves three primary parameters: ion column density, Doppler-$b$ parameter, and velocity shift of the line center. For each \osv or \oeit absorption feature, we fitted a single Voigt profile without attempting to separate the contributions of MW and M\,31. This choice is justified because the spectral resolution of RGS ($\sim70$~m\AA, corresponding to a velocity width of $\sim1000$~km\,s$^{-1}$ around $20$~\AA) is much larger than the LOS velocity separation between the two galaxies ($\sim100$~km\,s$^{-1}$, \citealp{van2008}). Consequently, the MW and M\,31 components are unresolved and blended. The ion column density was allowed to vary between $10^{15}$ and $10^{19}~\text{cm}^{-2}$, the velocity shift between $-750$ and $750$~km\,s$^{-1}$, and the Doppler-$b$ parameter between $20$ and $300$~km\,s$^{-1}$. The line equivalent width (EW) was then calculated by integrating the Voigt optical-depth profile using the best-fit ion column density, Doppler-$b$ parameter, and line-center velocity shift, following Section~3.1 of \citet{Buote2009}. Additional Voigt profiles were included when necessary to model other prominent line features in the segmental spectra, thereby providing better constraints on the continuum level. The best fits were obtained by minimizing the $C$-statistic (e.g., \citealp{Kaastra2017}). The fitting results and example \osv and \oeit spectra are presented in Appendix~\ref{append:fit_result}.

Due to the blending of the MW and M31 absorption, we search for excess absorption relative to the MW foreground to trace the hot CGM of M\,31. At large impact parameters, the measured absorption is dominated by the MW, which we initially approximate as constant across the M\,31 field. We test the stability of this empirical foreground estimate in Section~\ref{subsec:excess_absorption} and discuss possible spatial variation in Section~\ref{subsec:patchiness}. In contrast, inner sightlines should exhibit an absorption excess due to an additional contribution from M\,31.

\section{Results} 
\label{sec:results}

\subsection{Excess Absorption toward M\,31}
\label{subsec:excess_absorption}
\begin{figure}
\centering
\includegraphics[width=0.98\linewidth]{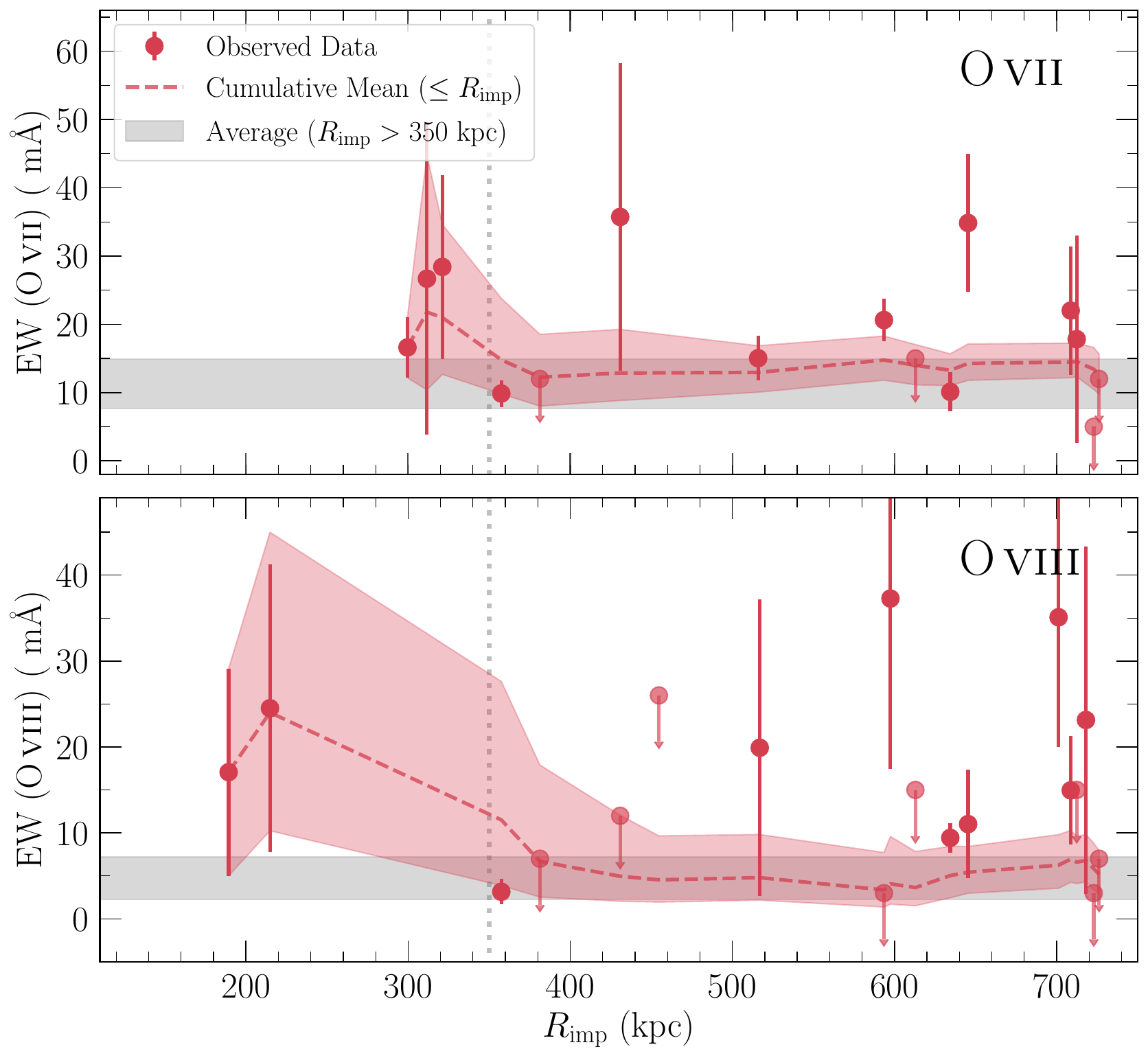}
\caption{Observed LOS total EW of $z\sim0$ \osv (top) and \oeit (bottom) lines as a function of impact parameter. The points represent the measurements, with error bars indicating $1\,\sigma$ uncertainties. The downward arrows mark the $3\,\sigma$ upper limits for non-detections. The dashed red curve and the corresponding shaded region represent the cumulative mean EW and its $1\,\sigma$ uncertainty, calculated at each point by averaging all measurements at smaller or equal impact parameters. The gray horizontal band is the average EW measurement for LOSs outside $R_{\rm imp}=350$~kpc (i.e., the vertical dotted line), which is consistent with the MW foreground absorption.
}
\label{fig:ew_r}
\end{figure}

Figure~\ref{fig:ew_r} shows the EWs of \osv and \oeit absorption lines as a function of impact parameter. The measurements are overlaid with the cumulative mean (dashed red line), computed at each data point using all measurements with smaller or equal impact parameters. The gray shaded band indicates the mean EW measured from sightlines outside $R_{\rm imp}=350$~kpc. All mean EWs are derived within a Bayesian framework that incorporates both detections and upper limits (Appendix~\ref{append:bayesian_method}).

For both \osv and O\,{\sc viii}, the cumulative mean EW increases toward the center of M\,31, aligning with the scenario of an extended hot halo surrounding M\,31. This inner excess is also consistent with our previous discovery toward PG\,0052+251. In that work, multiple absorption lines, including N\,{\sc vii}, O\,{\sc vii}, O\,{\sc viii}, Ne\,{\sc ix}, etc., were jointly modeled with a collisional ionization equilibrium (CIE) plasma model, revealing an excess $z\sim0$ hot-gas hydrogen column density beyond the expected MW foreground contribution \citep{Zhou2025}.

\begin{deluxetable*}{c|cllc|cllc|c}
\tablecaption{Average EWs and the Significance of Inner Excess with Different  Sky Boundaries \label{tab:exces_abs}}
\tabletypesize{\scriptsize}
\tablewidth{0pt}
\setlength{\tabcolsep}{2.8mm}
\tablehead{
Sky & \multicolumn{4}{c|}{\osv} & \multicolumn{4}{c|}{\oeit} & \\
\cline{2-9}
Boundary & \colhead{Num. LOS} & \colhead{${\rm \left < EW_{in} \right >}$} & \colhead{${\rm \left < EW_{out} \right >}$} & $S/N_{\rm exc}$ &  \colhead{Num. LOS} & \colhead{${\rm \left < EW_{in} \right >}$} & \colhead{${\rm \left < EW_{out} \right >}$} & $S/N_{\rm exc}$ & \colhead{$S/N_{\rm total}$} \\
(kpc) & \colhead{(in/out)} & \colhead{(m\AA)} & \colhead{(m\AA)} & ($\sigma$) & \colhead{(in/out)} & \colhead{(m\AA)} & \colhead{(m\AA)} & ($\sigma$) & \colhead{($\sigma$)} 
}
\startdata
$300$ & $1/14$ & $16.61\pm4.44$ & $12.25^{+3.50}_{-3.32}$ & $1.2$ & $2/16$ & $22.20^{+15.64}_{-12.28}$ & $4.56^{+2.65}_{-2.26}$ & $1.8$ & $2.2$ \\
$350$ & $3/12$ & $20.33^{+10.81}_{-7.82}$ & $11.34^{+3.53}_{-3.64}$ & $1.4$ & $2/16$ & $22.20^{+15.64}_{-12.28}$ & $4.56^{+2.65}_{-2.26}$ & $1.8$ & $2.3$ \\
$400$ & $5/10$ & $12.18^{+6.20}_{-4.25}$ & $12.68^{+4.76}_{-4.61}$ & $<1.0$ & $4/14$ & $6.02^{+9.24}_{-3.64}$ & $5.64^{+3.65}_{-3.10}$ & $<1.0$ & $<1.0$ \\
\enddata
\tablecomments{For each ion, Num. LOS gives the number of LOSs inside/outside the adopted sky boundary. ${\rm \left < EW_{in} \right >}$ and ${\rm \left < EW_{out} \right >}$ are the average EWs within and outside boundary. $S/N_{\rm exc}$ is the significance of the inner excess relative to the outer average. $S/N_{\rm total}$ combines the \osv\ and \oeit\ excess significances in quadrature.
}
\end{deluxetable*}

To quantify the excess absorption, we compared the mean \osv and \oeit EWs between the inner and outer regions around M\,31. Here, we tested three sky boundaries of $R_{\rm imp} = 300$, $350$, and $400$~kpc to divide the sample into inner and outer sightlines. The mean EWs in each region were calculated using the same Bayesian method. The results are summarized in Table~\ref{tab:exces_abs}.

The mean EWs in the outer region are approximately $12$~m\AA\ for \osv and $5$~m\AA\ for O\,{\sc viii}, and are stable against the choice of sky boundary. These values are consistent with the expected MW foreground absorption predicted from several Galactic hot-gas density models \citep{Li2017, Troitsky2017, Kaaret2020, Martynenko2022, Locatelli2024}, which give $11.8-19.2$~m\AA\ for \osv and $2.1-7.3$~m\AA\ for \oeit (see \citealp{Zhou2025} for details).

For plausible sky boundaries of $R_{\rm imp}=300$--$350$~kpc, the mean excess EW ranges from $4.28^{+5.66}_{-5.64}$ to $9.20^{+10.90}_{-8.99}$~m\AA\ for O\,{\sc vii}, whose inner sightlines lie at $R_{\rm imp}\sim300$--$320$~kpc. For O\,{\sc viii}, the two inner sightlines probe smaller impact parameters of
$R_{\rm imp}\sim190$--$215$~kpc and yield an excess of $17.43^{+15.83}_{-12.49}$~m\AA. The corresponding significance levels are $1.2$--$1.4\,\sigma$ for \osv and $1.8\,\sigma$ for O\,{\sc viii}, giving a nominal combined significance of $2.2$--$2.3\,\sigma$. When the boundary is extended to $400$~kpc, the excess becomes statistically insignificant, indicating that the signal is driven primarily by the innermost sightlines.

The significance quoted above is derived from the comparison with the empirically estimated foreground, following standard approaches adopted in previous X-ray absorption studies. While the methodology is robust, the limited number of inner sightlines and the modest signal-to-noise ratio lead us to interpret the result as a marginal excess rather than a firm detection.

\subsection{Mass Implications for the Hot CGM of M\,31}
The virial mass of M\,31 constrained by previous studies typically ranges from $(0.7-3.0)\times10^{12}~M_\odot$, with a most likely value of $\sim1.6\times10^{12}~M_\odot$ \citep{Bhattacharya2025}. Adopting the cosmic baryon fraction, $f_b = \Omega_b/\Omega_m\simeq15.7\%$ \citep{Planck2020}, the expected baryon mass of M\,31 is $(1.10-4.71)\times10^{11}~M_\odot$. The currently known baryon amount in M\,31 is approximately $1.45\times10^{11}~M_\odot$, including stars ($1.0\times10^{11}~M_\odot$, \citealt{Seigar2008, Williams2017}), cold disk gas ($8\times10^9~M_\odot$, \citealt{Corbelli2010}), and cool to warm phase CGM ($3.7\times10^{10}~M_\odot$, \citealt{Lehner2020}). This leaves a possible ``missing'' baryon mass of up to $3.26\times10^{11}~M_\odot$, with the most likely value being $1.06\times10^{11}~M_\odot$, before accounting for the hot gas.

We estimated the hot-CGM mass in M\,31 implied by the observed excess absorption by fitting a modified-$\beta$ halo profile \citep{Cavaliere1976, Miller2013, Miller2015}. The hydrogen density of hot CGM is described as $n_{\rm H}(r) = C_\beta r^{-3\beta}$, where $r$ is the galactocentric radius in kpc, $C_\beta$ is the density normalization in cm$^{-3}$\,kpc$^{3\beta}$, and $\beta$ defines the radial slope of the density decline. 

Due to the limited significance of the excess, the current data cannot simultaneously constrain all parameters related to the $\beta$-model fitting. Therefore, we fixed $\beta=0.4$ based on a recent {\it eROSITA} study of M\,31-mass galaxies \citep{Zhang2024}. This value is broadly consistent with observational constraints for massive spiral and $L^*$ galaxies ($\beta\sim0.35-0.6$, e.g., \citealp{Li2017ApJS, Singh2018, Bregman2018, Bregman2022, He2026arXiv}). Although it is somewhat flatter than the hot-gas density slopes found for star-forming $L^*$ halos in TNG100 ($\beta\sim0.5$--$0.6$; \citealt{Oren2024}), flatter slopes of $\beta\sim0.2$--$0.35$ have also been reported in analytic models that include metallicity gradients (e.g., \citealt{Faerman2020, Martynenko2022}). A gas metallicity of $0.3~Z_\odot$ \citep{Lodders2009} and a temperature of $2.5\times10^6$~K were adopted to calculate ion fractions under CIE \citep{Gnat2007}. These values are consistent with previous measurements for the MW-like galaxies \citep{Li2017, Troitsky2017, Bregman2018, Kaaret2020, Martynenko2022, Bregman2022}. 

We integrated along LOSs for \osv and \oeit column densities contributed by M\,31. Only gas within an assumed CGM boundary, $R_{\rm CGM}$, was considered, which was tested from $300-500$~kpc. The model column densities were then converted to line EWs using the curve-of-growth method, assuming a Doppler-$b$ parameter of $120$~km\,s$^{-1}$ \citep{Gupta2012, Fang2015, Nicastro2016, Li2017}. Details of the model calculation and fitting procedure are given in Appendix~\ref{append:beta_model}. 

Figure~\ref{fig:mhalo_rcgm} shows the best-fit hot CGM mass of M\,31 as a function of the assumed CGM boundary. The inverse relationship between the inferred mass and $R_{\rm CGM}$ arises because, for fixed $\beta$, the gas mass enclosed within the fixed $R_{\rm vir}$ is proportional to the fitted density normalization, $C_\beta$. A smaller $R_{\rm CGM}$ shortens the LOS path length through the M\,31 CGM and therefore requires a larger $C_\beta$ to reproduce the observed absorption attributed to M\,31, producing a larger inferred CGM mass. If all excess absorption arises from gas confined within $R_{\rm CGM}\lesssim380$~kpc, the inferred hot CGM mass becomes implausibly large, exceeding the possible range of the ``missing'' mass of M\,31. For adopted CGM boundaries of $400$--$500$~kpc, the inferred hot-gas mass enclosed within $R_{\rm vir}$ decreases from $2.1^{+2.4}_{-1.3}\times10^{11}~M_\odot$ to $1.1^{+1.4}_{-0.7}\times10^{11}~M_\odot$, while the total mass enclosed within $R_{\rm CGM}$ decreases from $3.4^{+4.0}_{-2.2}\times10^{11}~M_\odot$ to $2.7^{+3.4}_{-1.8}\times10^{11}~M_\odot$.

\begin{figure}
\centering
\includegraphics[width=0.98\linewidth]{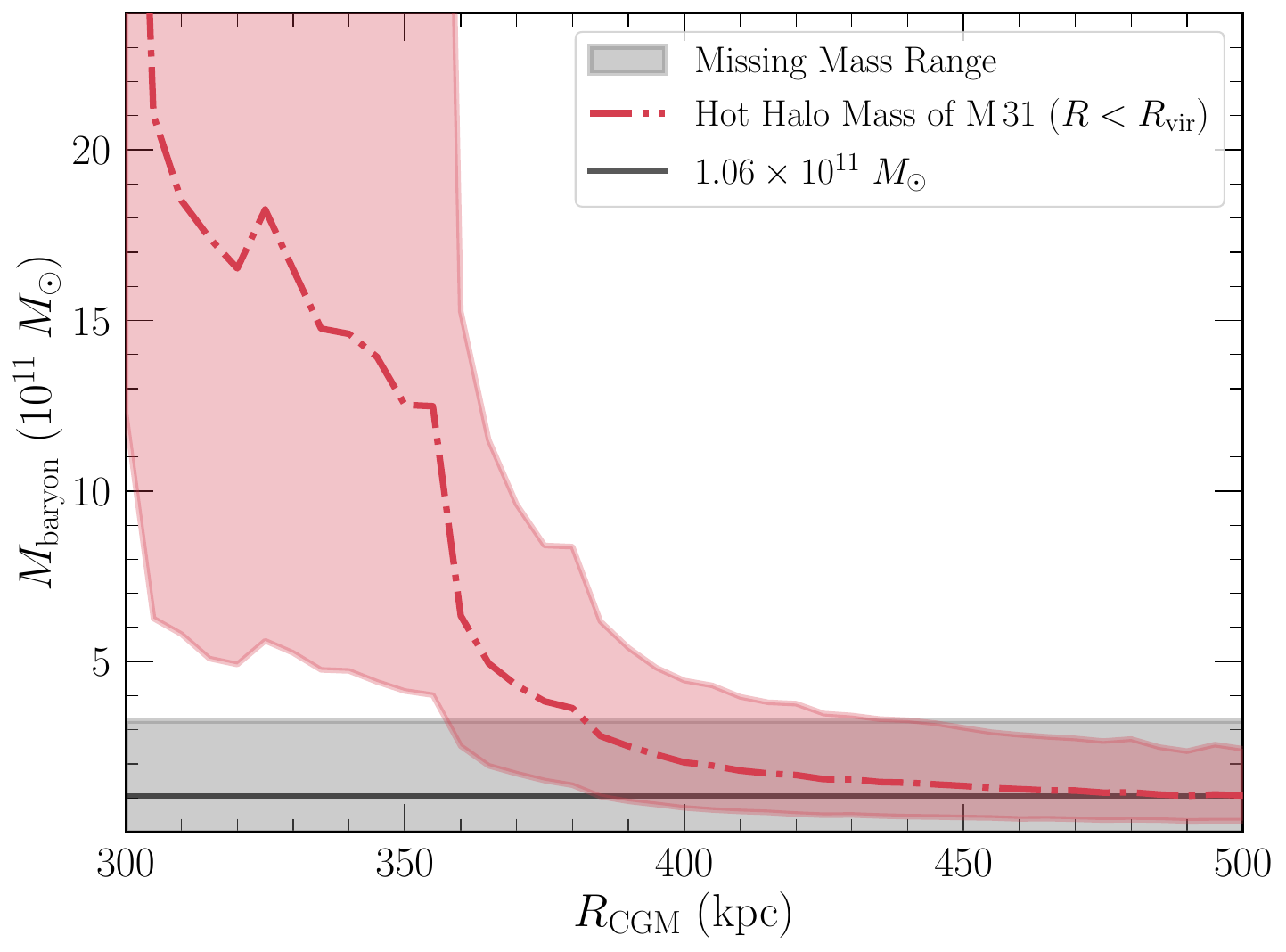}
\caption{Hot-CGM mass of M\,31 enclosed within $R_{\rm vir}$ as a function of the adopted CGM boundary, $R_{\rm CGM}$. The red dash-dotted curve and red shaded region represent the best-fit mass and its $1\,\sigma$ uncertainty. The black solid line and the gray shadow are the most likely value and the possible range of the ``missing'' baryon mass of M\,31. 
}
\label{fig:mhalo_rcgm}
\end{figure}

The mass estimate is highly uncertain and depends sensitively on the assumed CGM boundary, metallicity, temperature, and density profile (see also \citealp{Zhang2026}). Varying the fixed $\beta$ value or the gas temperature can change the inferred CGM mass by factors of a few, but the very large mass inferred when $R_{\rm CGM}$ is close to $R_{\rm vir}$ remains difficult to accommodate. At present, the data do not provide a robust constraint on the baryon content of the M\,31's hot halo. The inferred mass exceeds the nominal ``missing'' baryon budget under some assumptions, further underscoring the large systematic uncertainties in this conversion.

\section{Discussion} 
\label{sec:discussion}

The observed excess absorption is consistent with the expected scenario of an extended hot halo surrounding M\,31. However, the derived large $R_{\rm CGM}$ for a reasonable mass range does not necessarily mean that the hot CGM of M\,31 is extended. Possible sources along LOSs may contribute to this excess and complicate the analysis of gas distribution. Below, we present a discussion on these possibilities.

\subsection{Foreground Patchiness}
\label{subsec:patchiness}
The first caveat in our analysis is the patchiness of the MW foreground. X-ray emission studies have shown that the temperature and emission measure of the Galactic hot gas can vary substantially even between nearby LOSs (e.g., \citealt{Yoshino2009, Henley2010, Nakashima2018, Kaaret2020}). This inhomogeneity complicates the foreground assessment and may either enhance or suppress the excess absorption observed toward M\,31. However, emission-derived patchiness likely overestimates the corresponding variation in absorption, because absorption scales linearly with gas density, whereas emission scales approximately as density squared and can also be affected by foreground neutral absorption.

Recent emission studies found an autocorrelation scale of $\sim5$--$10^{\circ}$ for this patchiness, possibly linked to stellar feedback \citep{Kaaret2020, Qu2024}. In addition, {\it eROSITA} observations of the eFEDS field revealed that variations of the hot gas emission are strongly anti-correlated with the column density of Galactic neutral gas \citep{Ponti2023_patchiness}. Their discovery suggests that the intrinsic scatter of the hot halo may be $\lesssim25\%$ after accounting for the absorption effects of the disk neutral gas. If this result is representative of the entire sky, and if the Galactic hot-gas absorption arises mainly from the hot halo instead of the disk gas toward high Galactic latitudes \citep{Kaaret2020, Locatelli2024}, a $25\%$ scatter would introduce foreground uncertainties of $\lesssim4$~m\AA\ for \osv and $\lesssim1.5$~m\AA\ for O\,{\sc viii}, which has a minor impact on our results.

\subsection{Intra-group Medium}
\label{subsec:IGrm}
The second caveat is that part of the observed excess may arise from the intra-group medium. A recent study using $171$ fast radio bursts (FRBs) toward M\,31 reported a closure radius (i.e., the characteristic radius at which the enclosed baryon fraction approaches the cosmic mean) of $9.2^{+9.9}_{-4.9}~R_{\rm vir}$ for the ionized baryon halo surrounding M\,31 \citep{Kahinga2026}. This result suggests that a substantial amount of ionized gas may extend well beyond the virial radius.

Additionally, using X-ray emission and the thermal Sunyaev-Zel'dovich (SZ) effect, \cite{Qu2021} found that the M\,31's hot halo alone can barely explain the enhanced hot-gas signals they detected toward M\,31. Otherwise, the inferred CGM mass would be comparable to the total mass of the Local Group. This result is qualitatively similar to our discovery via X-ray absorption, in which attributing all the excess absorption to hot CGM within the virial radius of M\,31 would require an unreasonably large hot-gas mass. 

To explain the strong signal, \cite{Qu2021} proposed a hot-gas bridge connecting the MW and
M\,31 as an additional contributor. This scenario was later supported by a study using two FRBs that intersect the inner halo of M\,31 \citep{Anna-Thomas2025}. Adopting the fiducial-SZ model of \cite{Qu2021} (see their Table~1 for details), we estimate a bridge contribution of ${\rm EW_{\rm OVII,\,bri}}\sim2.5$~m\AA\ toward our \osv LOSs with $R_{\rm imp}\sim300$--$320$~kpc, and ${\rm EW_{\rm OVIII,\,bri}}\sim3$~m\AA\ toward the two innermost \oeit LOSs. These estimates are insensitive to the assumed Doppler-$b$ parameter, as the predicted bridge absorption is weak and remains close to the linear regime of the curve of growth.

\subsection{Photoionization from Metagalactic Background}
\label{subsec:photoionization}
Photoionization by the cosmic ultraviolet and X-ray background may affect the ionization balance of \osv and \oeit in the CGM (e.g., \citealp{Oppenheimer2013, Nelson2018, Wijers2019, Faerman2020}). Our modeling of gas distribution and total CGM mass assumes a CIE condition, which may introduce systematic uncertainties if the metagalactic radiation field contributes significantly to the ionization of the low-density CGM at large radii.

This effect is expected to be stronger at lower gas densities and temperatures. In particular, \oeit is more sensitive to photoionization than O\,{\sc vii}, because the collisional \osv fraction remains high over a broad temperature range ($\sim3\times10^5$--$2\times10^6$~K), whereas the \oeit fraction drops rapidly below $\sim2\times10^6$~K (e.g., \citealt{Gnat2007}). Photoionization may significantly modify the \osv and \oeit fractions when CGM density lies below a few $10^{-5}-10^{-4}$~cm$^{-3}$ \citep{Nelson2018, Faerman2020}, as expected for hot CGM at a few hundred kpc. However, the magnitude of this effect remains uncertain and depends on several assumptions, such as the radiation field, gas density profile, temperature structure, and gas morphology. A detailed treatment of photoionization is therefore beyond the scope of this paper.

\section{Summary} 
\label{sec:Summary}
We present the first \osv and \oeit absorption-line survey of the hot CGM of M\,31 using {\it XMM-Newton}/RGS. A total of $15$ and $18$ background AGNs are selected to compile the \osv and \oeit samples, respectively, covering impact parameters of approximately $R_{\rm imp}\sim300$--$730$~kpc and $\sim190$--$730$~kpc.

We measure the LOS total \osv and \oeit absorption at zero redshift and compare the results with the expected MW foreground. Comparing sightlines inside and outside $R_{\rm imp}=300$--$350$~kpc, we find a marginal excess absorption toward the inner region of M\,31. The excess EWs are $\sim4-9$~m\AA\ and $\sim17$~m\AA\ for our innermost \osv and \oeit sightlines at $R_{\rm imp}\sim310$~kpc and $\sim200$~kpc, respectively (see Table~\ref{tab:exces_abs} for details). The nominal combined significance for this excess is $2.2$--$2.3\,\sigma$. When the sky boundary is extended to $400$~kpc, the excess becomes statistically insignificant. The mean \osv and \oeit EWs in the corresponding outer regions are consistent with the expected MW contribution.

Attributing this excess entirely to the hot CGM of M\,31, the inferred hot-gas mass depends strongly on the adopted CGM boundary, $R_{\rm CGM}$. It would exceed the nominal ``missing'' baryon budget for $R_{\rm CGM} \lesssim380$~kpc, and is of order a few $\times10^{11}~M_\odot$ for larger CGM boundaries. This mass estimate remains highly uncertain and is sensitive to several model assumptions, yet it still implies that the hot CGM constitutes a substantial reservoir of baryons.

\begin{acknowledgments}
Z.Z. and T.F. acknowledge support from the National SKA Program of China No.\ 2025SKA0150103, National Natural Science Foundation of China under Nos.\ 12550002, 12133008, 12221003, 11890692, and the science research grant from the China Manned Space Project with No.\ CMS-CSST-2021-A04 and No. CMS-CSST-2025-A10. F.N. acknowledges support from the INAF-PRIN grant ``A Systematic Study of the Largest Reservoir of Baryons and Metals in the Universe: the Circum-Galactic Medium of Galaxies'' (No.\ 1.05.01.85.10), the EU
HORIZON-2020 grant ``AHEAD2020'' (Agreement No.\ 871158), and the INAF-AF-2023 project ``The XRISM-to-XIFU (X2X) Agreement and Beyond: Entering a New Era of High-Resolution X-Ray Spectroscopy'' (No.\ 1.05.23.01.06).
\end{acknowledgments}

\bibliography{ref}{}
\bibliographystyle{aasjournalv7}

\appendix
\section{Spectral Fitting Result}
\label{append:fit_result}

\begin{figure*}
\centering
\includegraphics[width=0.95\linewidth]{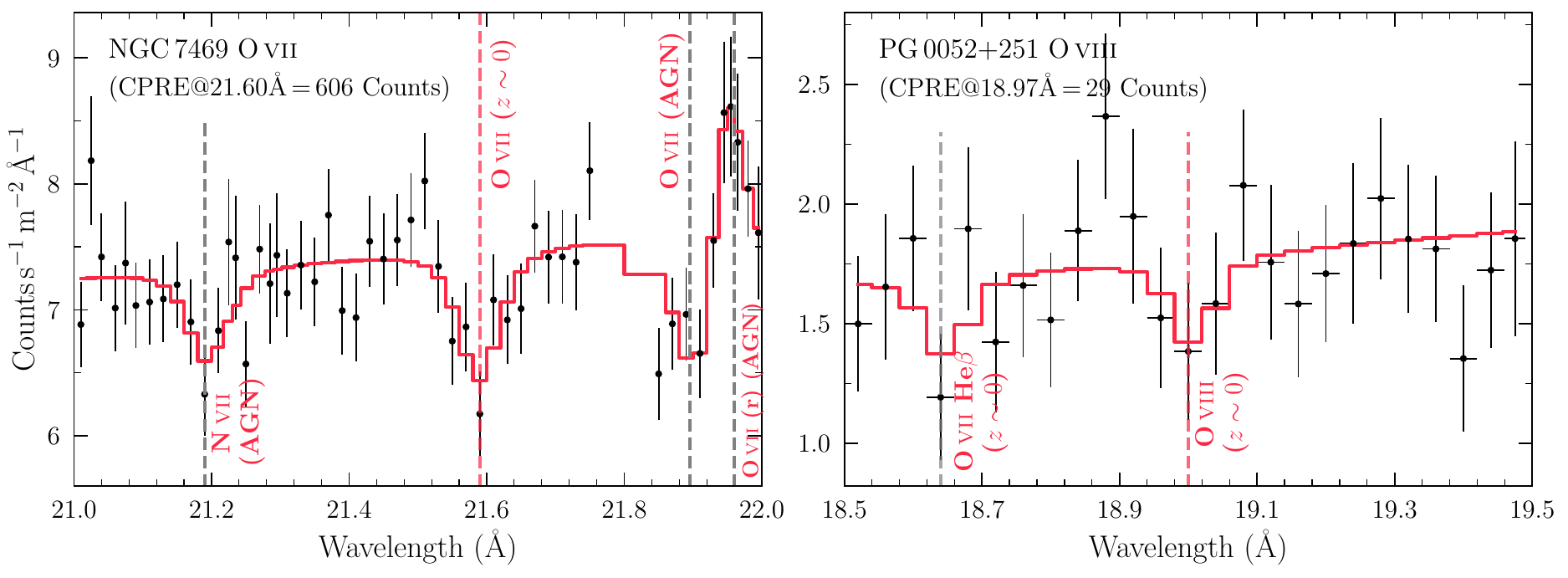}
\caption{Stacked \osv spectrum of NGC\,7469 (left) and \oeit spectrum of PG\,0052+251 (right). \osv and \oeit K$\alpha$ lines at $z\sim0$ are highlighted by vertical red lines. The left panel illustrates how intrinsic line features associated with the AGN \citep{Behar2017} are treated in our analysis.}
\label{fig:example_spec}
\end{figure*}

Tables~\ref{tab:ovii} and \ref{tab:oviii} summarize the properties of the selected AGN sources and the spectral fitting results for the \osv and \oeit samples, respectively. The detailed spectral fitting method is described in section~\ref{subsec:spec_analysis}. The XSPEC command, ``{\it error}'', was used to calculate the uncertainties of the fitting parameters, while the uncertainties in EW were estimated through $1,000$ Monte Carlo simulations following the method of \citet{Luo2018}. All measurement uncertainties are quoted at $1\,\sigma$ significance range. As for non-detections, we determined the $3\,\sigma$ upper limits of EW based on the spectral CPRE (see, \citealp{Fang2005}). The EW upper limits were later converted to upper limits of ion column density using the curve-of-growth technique, assuming a Doppler-$b$ parameter of $120$~km\,s$^{-1}$~\citep{Li2017}. The last column of each table lists the references we adopted to identify the AGN intrinsic lines in the $z\sim0$ \osv\ and \oeit\ spectral segments, which were also modeled in our spectral fitting procedure using additional Voigt profiles. Example \osv and \oeit spectra are shown in Figure~\ref{fig:example_spec}.

\begin{deluxetable}{lccrrcccccccrr}[ht]
\tablecaption{Best-fit Result for the \osv Sample \label{tab:ovii}}
\tabletypesize{\scriptsize}
\tablewidth{0pt}
\setlength{\tabcolsep}{0.95mm}
\tablehead{
\colhead{Name} & \colhead{Type} & \colhead{$z$} & \colhead{$l$} & \colhead{$b$} &  \colhead{$R_{\rm imp}$}  & \colhead{CPRE} & \colhead{$\log N$ (O\,{\sc vii})} & \colhead{Doppler-$b$} & \colhead{$v_{\rm shift}$} &
\colhead{EW (O\,{\sc vii})} & \colhead{S/N} & \colhead{$C_{\rm stat} / {\rm DoF}$} & \colhead{Ref.} \\
\colhead{} & \colhead{} & \colhead{} & \colhead{(deg)} & \colhead{(deg)} & \colhead{(kpc)} & \colhead{(@$21.60$~\AA)} & \colhead{(cm$^{-2}$)} & \colhead{(km\,s$^{-1}$)} & \colhead{(km\,s$^{-1}$)} &
\colhead{(m\AA)} & \colhead{($\sigma$)} & \colhead{} \\
\colhead{(1)} & \colhead{(2)} & \colhead{(3)} & \colhead{(4)} & \colhead{(5)} & \colhead{(6)} & \colhead{(7)} & \colhead{(8)} & \colhead{(9)} & \colhead{(10)} &
\colhead{(11)} & \colhead{(12)} & \colhead{(13)} & \colhead{(14)}
}
\startdata
Mrk\,335 & 2 & $0.0259$ & $108.763$ & $-41.424$ & $299.6$ & $364$ & $16.33^{+1.55}_{-0.56}$ & $74^{+192}_{-54}$ & $-247^{+120}_{-124}$ & $16.61\pm4.44$ & $3.7$ & $106/92$ & [1] \\
Mrk\,1040 & 2 & $0.0163$ & $146.114$ & $-27.166$ & $311.6$ & $26$ & $16.55^{+2.08}_{-1.11}$ & $121^{+179}_{-101}$ & $-629^{+479}_{-121}$ & $26.69\pm22.86$ & $1.2$ & $84/92$ & [2] \\
Mrk\,359 & 2 & $0.0168$ & $134.596$ & $-42.874$ & $321.1$ & $45$ & $16.14^{+2.57}_{-0.49}$ & $300^{+0}_{-280}$ & $-750^{+347}_{-0}$ & $28.41\pm13.49$ & $2.1$ & $96/92$ & [3] \\
Ark\,564 & 2 & $0.0243$ & $92.138$ & $-25.337$ & $357.5$ & $1567$ & $16.05^{+1.29}_{-0.52}$ & $46^{+194}_{-26}$ & $-17^{+91}_{-107}$ & $9.86\pm1.97$ & $5.0$ & $97/92$ & [4-6] \\
Mrk\,1502 & 2 & $0.0612$ & $123.749$ & $-50.175$ & $381.3$ & $124$ & $<15.8$ & \nodata & \nodata & $<12$ & \nodata & $114/92$ & [7, 8] \\
1H\,0323+342 & 1 & $0.0626$ & $155.727$ & $-18.757$ & $430.8$ & $37$ & $17.13^{+1.73}_{-1.28}$ & $128^{+172}_{-108}$ & $-457^{+260}_{-252}$ & $35.73\pm22.56$ & $1.6$ & $89/92$ & \nodata \\
NGC\,7469 & 2 & $0.0160$ & $83.098$ & $-45.467$ & $515.9$ & $606$ & $16.28^{+1.55}_{-0.57}$ & $68^{+232}_{-48}$ & $-142^{+121}_{-117}$ & $15.03\pm3.29$ & $4.6$ & $75/92$ & [9] \\
1ES\,1959+650 & 1 & $0.0470$ & $98.004$ & $17.670$ & $593.5$ & $724$ & $16.56^{+1.32}_{-0.58}$ & $86^{+130}_{-51}$ & $-92^{+74}_{-80}$ & $20.63\pm3.08$ & $6.7$ & $96/92$ & \nodata \\
Mrk\,590 & 2 & $0.0265$ & $163.500$ & $-56.942$ & $612.9$ & $75$ & $<16.0$ & \nodata & \nodata & $<15$ & \nodata & $95/92$ & [10, 11] \\
1ES\,1927+654 & 2 & $0.0170$ & $96.984$ & $20.961$ & $634.4$ & $1017$ & $15.96^{+1.48}_{-0.51}$ & $53^{+179}_{-33}$ & $-33^{+113}_{-114}$ & $10.11\pm2.86$ & $3.5$ & $95/92$ & [12] \\
3C\,390.3 & 2 & $0.0556$ & $111.437$ & $27.074$ & $645.4$ & $58$ & $17.01^{+1.60}_{-0.94}$ & $121^{+179}_{-101}$ & $589^{+158}_{-169}$ & $34.84\pm10.10$ & $3.4$ & $100/92$ & [13] \\
Mrk\,926 & 2 & $0.0477$ & $64.091$ & $-58.756$ & $708.7$ & $66$ & $16.96^{+1.44}_{-1.22}$ & $75^{+225}_{-55}$ & $718^{+32}_{-273}$ & $22.01\pm9.38$ & $2.3$ & $100/92$ & \nodata \\
PG\,0804+761 & 2 & $0.0988$ & $138.279$ & $31.033$ & $712.4$ & $32$ & $16.41^{+1.82}_{-1.38}$ & $78^{+217}_{-63}$ & $476^{+274}_{-333}$ & $17.82\pm15.18$ & $1.2$ & $90/92$ & \nodata \\
Mrk\,1044 & 2 & $0.0173$ & $179.694$ & $-60.477$ & $722.9$ & $880$ & $<15.3$ & \nodata & \nodata & $<5$ & \nodata & $84/92$ & [14] \\
NGC\,985 & 2 & $0.0430$ & $180.837$ & $-59.490$ & $726.2$ & $125$ & $<15.8$ & \nodata & \nodata & $<12$ & \nodata & $103/92$ & [15] \\
\enddata
\tablecomments{Columns are: (1) target name; (2) target type, where 1 and 2 denote BL Lac and Seyfert, respectively, with BL Lacs being free from intrinsic AGN contamination; (3) target redshift; (4-5) Galactic coordinates; (6) impact parameter; (7) CPRE at around $21.60$~\AA; (8-11) best-fit column density, Doppler-$b$ parameter, velocity shift, and equivalent width of the \osv K$\alpha$ line; (12) signal-to-noise ratio of the detected \osv line; (13) $C_{\rm stat}$ over degrees of freedom for the best-fit model. Targets are sequenced by increasing impact parameter. Column~(14) lists the references we adopted for identifying the wavelengths of the AGN intrinsic lines, which are: [1] \cite{Longinotti2019}; [2] \cite{Reeves2017}; [3] \cite{O'Brien2001}; [4] \cite{Smith2008}; [5] \cite{Gupta2013}; [6] \cite{Khanna2016}; [7] \cite{Silva2018}; [8] \cite{Reeves2019}; [9] \cite{Behar2017}; [10] \cite{Longinotti2007}; [11] \cite{Gupta2015}; [12] \cite{Ricci2021}; [13] \cite{Tombesi2016}; [14] \cite{Krongold2021}; [15] \cite{Ebrero2021}. ``...'' indicates that no X-ray AGN wind has been previously reported.
}
\end{deluxetable}
\onecolumngrid

\begin{deluxetable}{lccrrcccccccrr}
\tablecaption{Best-fit Result for the \oeit Sample \label{tab:oviii}}
\tabletypesize{\scriptsize}
\tablewidth{0pt}
\setlength{\tabcolsep}{0.95mm}
\tablehead{
\colhead{Name} & \colhead{Type} & \colhead{$z$} & \colhead{$l$} & \colhead{$b$} &  \colhead{$R_{\rm imp}$}  & \colhead{CPRE} & \colhead{$\log N$ (O\,{\sc viii})} & \colhead{Doppler-$b$} & \colhead{$v_{\rm shift}$} &
\colhead{EW (O\,{\sc viii})} & \colhead{S/N} & \colhead{$C_{\rm stat} / {\rm DoF}$} & \colhead{Ref.}  \\
\colhead{} & \colhead{} & \colhead{} & \colhead{(deg)} & \colhead{(deg)} & \colhead{(kpc)} & \colhead{(@$18.97$~\AA)} & \colhead{(cm$^{-2}$)} & \colhead{(km\,s$^{-1}$)} & \colhead{(km\,s$^{-1}$)} &
\colhead{(m\AA)} & \colhead{($\sigma$)} & \colhead{} \\
\colhead{(1)} & \colhead{(2)} & \colhead{(3)} & \colhead{(4)} & \colhead{(5)} & \colhead{(6)} & \colhead{(7)} & \colhead{(8)} & \colhead{(9)} & \colhead{(10)} &
\colhead{(11)} & \colhead{(12)} & \colhead{(13)} & \colhead{(14)}
}
\startdata
1ES\,2344+514 & 1 & $0.0443$ & $112.891$ & $-9.908$ & $189.3$ & $57$  & $16.86^{+1.86}_{-1.20}$ & $80^{+220}_{-60}$ & $689^{+61}_{-370}$ & $17.07\pm12.06$ & $1.4$ & $92/99$ & \nodata \\
PG\,0052+251 & 2 & $0.1551$ & $123.907$ & $-37.437$ & $214.8$ & $29$ & $17.09^{+1.91}_{-1.28}$ & $110^{+190}_{-90}$ & $628^{+122}_{-458}$ & $24.51\pm16.70$ & $1.5$ & $104/99$ & [1] \\
Ark\,564 & 2 & $0.0243$ & $92.138$ & $-25.337$ & $357.5$ & $4061$ & $15.50^{+0.75}_{-0.36}$ & $28^{+272}_{-8}$ & $13^{+350}_{-273}$ & $3.16\pm1.48$ & $2.1$ & $104/99$ & [2-4] \\
Mrk\,1502 & 2 & $0.0612$ & $123.749$ & $-50.175$ & $381.3$ & $323$ & $<15.9$ & \nodata & \nodata & $<7$ & \nodata & $100/99$ & [5, 6] \\
1H\,0323+342 & 1 & $0.0626$ & $155.727$ & $-18.757$ & $430.8$ & $127$ & $<16.2$ & \nodata & \nodata & $<12$ & \nodata & $98/99$ & \nodata \\
3C\,454.3 & 1 & $0.8579$ & $86.111$ & $-38.184$ & $454.7$ & $25$ & $<17.1$ & \nodata & \nodata & $<26$ & \nodata & $99/99$ & \nodata \\
PG\,2209+184 & 1 & $0.0696$ & $78.368$ & $-30.047$ & $516.9$ & $34$ & $17.10^{+1.90}_{-1.58}$ & $85^{+215}_{-65}$ & $407^{+343}_{-402}$ & $19.91\pm17.28$ & $1.2$ & $103/99$ & \nodata \\
1ES\,1959+650 & 1 & $0.0470$ & $98.004$ & $17.670$ & $593.5$ & $2424$ & $<15.4$ & \nodata & \nodata & $<3$ & \nodata & $97/99$ & \nodata \\
Mrk\,1018 & 2 & $0.0426$ & $159.819$ & $-57.702$ & $597.5$ & $29$ & $16.70^{+2.30}_{-0.51}$ & $300^{+0}_{-280}$ & $750^{+0}_{-469}$ & $37.27\pm19.86$ & $1.9$ &  $100/99$ & \nodata \\
Mrk\,590 & 2 & $0.0265$ & $163.500$ & $-56.942$ & $612.9$ & $74$ & $<16.4$ & \nodata & \nodata & $<15$ & \nodata & $92/99$ & [7, 8] \\
1ES\,1927+654 & 2 & $0.0170$ & $96.984$ & $20.961$ & $634.4$ & $3007$ & $16.32^{+1.52}_{-0.46}$ & $54^{+194}_{-34}$ & $-168^{+109}_{-88}$ & $9.43\pm1.74$ & $5.4$ & $86/99$ & [9] \\
3C\,390.3 & 2 & $0.0556$ & $111.437$ & $27.074$ & $645.4$ & $160$ & $16.51^{+1.87}_{-0.94}$ & $56^{+244}_{-36}$ & $47^{+433}_{-294}$ & $11.03\pm6.30$ & $1.8$ & $102/99$ & [10] \\
MCG+08-11-11 & 2 & $0.0202$ & $165.731$ & $10.407$ & $701.2$ & $27$ & $17.42^{+1.58}_{-1.04}$ & $146^{+154}_{-126}$ & $345^{+219}_{-217}$ & $35.09\pm15.12$ & $2.3$ & $81/99$ & \nodata \\
Mrk\,926 & 2 & $0.0477$ & $64.091$ & $-58.758$ & $708.7$ & $165$ & $16.75^{+1.74}_{-0.88}$ & $73^{+227}_{-53}$ & $-303^{+217}_{-243}$ & $14.95\pm6.31$ & $2.4$  & $75/99$ & \nodata \\
PG\,0804+761 & 2 & $0.0988$ & $138.279$ & $31.033$ & $712.4$ & $75$ & $<16.4$ & \nodata & \nodata & $<15$ & \nodata & $92/99$ & \nodata \\
E\,1821+643 & 2 & $0.2970$ & $94.002$ & $27.417$  & $718.1$ & $32$ & $16.37^{+2.63}_{-0.95}$ & $300$ & $750^{+0}_{-619}$ & $23.14\pm20.21$ & $1.1$ & $97/99$ & \nodata \\
Mrk\,1044 & 2 & $0.0173$ & $179.694$ & $-60.477$ & $722.9$ & $2183$ & $<15.4$ & \nodata & \nodata & $<3$ & \nodata & $128/99$ & [11] \\
NGC\,985 & 2 & $0.0430$ & $180.837$ & $-59.490$ & $726.2$ & $291$ & $<15.9$ & \nodata & \nodata & $<7$ & \nodata & $115/99$ & [12] \\
\enddata
\tablecomments{The columns in this table are identical to those in Table \ref{tab:ovii} but for the \oeit sample. The references for AGN intrinsic line identification are: [1] \cite{Zhou2025}; [2] \cite{Smith2008}; [3] \cite{Gupta2013}; [4] \cite{Khanna2016}; [5] \cite{Silva2018}; [6] \cite{Reeves2019}; [7] \cite{Longinotti2007}; [8] \cite{Gupta2015}; [9] \cite{Ricci2021}; [10] \cite{Tombesi2016}; [11] \cite{Krongold2021}; [12] \cite{Ebrero2021}.
}
\end{deluxetable}
\onecolumngrid

\section{The Adopted Bayesian Framework}
\label{append:bayesian_method}
To incorporate upper limits into both the calculation of mean EWs and the fitting of the gas distribution, we adopted the Bayesian framework introduced in \cite{Lehner2026}. For detections, the standard Gaussian likelihood function was adopted:
\begin{equation}
\ln p(D_{\rm d}|\theta) = -\frac{1}{2}\sum_i[\frac{({\rm EW}_i - {\rm EW}_{{\rm model},\,i})^2}{\sigma_i^2} + \ln(2\pi\sigma_i^2)].    
\label{equ:likelihood_det}
\end{equation}
In the equation, $D_{\rm d}\equiv\{{\rm EW},\,\sigma_m\}$ represents the data for detections, where ${\rm EW}_i$ and $\sigma_{{\rm m},\,i}$ are the measured EW and its $1\sigma$ uncertainty for the $i$-th LOS, respectively. ${\rm EW_{\rm model}}$ is the model-predicted EW. When calculating the mean EW, ${\rm EW}_{{\rm model},\,i}$ is taken to be a constant for all LOSs.
$\sigma_i^2 = \sigma_{m,\,i}^2+ \sigma_p^2$ is the total uncertainty of the $i$-th LOS, where $\sigma_p$ is an additional parameter to account for intrinsic scatter beyond the measurement uncertainties. $\sigma_p$ is also known as the patchiness parameter, which has been used frequently while modeling the distribution of CGM (e.g., \citealp{Zheng2019, Kaaret2020, Qu2024, Lehner2026}). The parameter set $\theta$ consists of the mean EW and $\sigma_p$ for the average-EW estimation, and of $C_\beta$, the foreground EW, and $\sigma_p$ for the gas-distribution fitting.

For upper limits, the Gaussian cumulative distribution function (CDF) was used to express the probability that the true EW lies below a given upper limit:
\begin{equation}
\ln p(D_{\rm nd}|\theta) = \sum_i \ln \Phi(\frac{{\rm EW}_{{\rm upper},\,i} - {\rm EW}_{{\rm model},\,i}}{\sqrt{\sigma_{{\rm EW},i}^2 + \sigma_p^2}}).
\label{equ:likelihood_ndet}
\end{equation}
In the equation, $D_{\rm nd}\equiv\{{\rm EW_{upper}}\}$ denotes the observed data for non-detections, with ${\rm EW_{upper}}$ the $3\,\sigma$ upper limits of EW. $\sigma_{\rm EW}$ is the intrinsic $1\,\sigma$ EW uncertainty set by the spectral quality, regardless of line detection. Here, $\sigma_{\rm EW} = {\rm EW_{upper}}/3$. $\Phi$ is the standard CDF of the Gaussian distribution.

A uniform prior, $p(\theta)$, was adopted for the fitting parameters, and the posterior is written as:
\begin{equation}
p(\theta|D_{\rm all}) = p(D_{ \rm all}|\theta)p(\theta) = p(D_{\rm d}|\theta) p(D_{\rm nd}|\theta)p(\theta) ,
\label{equ:posterior}
\end{equation}
where $D_{\rm all} \equiv \{D_{\rm d},\,D_{\rm nd}\}$ is the observed data for both detections and non-detections. We sampled the posterior distribution using the Markov Chain Monte Carlo package, {\it emcee} \citep{Foreman-Mackey2013}. For each calculation, we implemented $32$ walkers for $5500$ steps, and the initial $500$ steps were discarded as burn-in. The median of the posterior distribution was adopted as the best-fit value, and the $16$th--$84$th percentile range was taken as the $1\sigma$ uncertainty.

\section{Modeling the Gas Distribution of the M\,31 Hot CGM}
\label{append:beta_model}
To model the hot-CGM distribution of M\,31, we expressed the observed EW along a given sightline as the sum of an M\,31 component and a foreground component:
\begin{equation}
{\rm EW_{ion}} (R_{\rm imp}) = {\rm EW_{ion,\,M31}} (R_{\rm imp}) + {\rm EW_{ion,\,fore}},   
\label{equ:ew_model}
\end{equation}
where ``ion'' represents either \osv or O\,{\sc viii}. For simplicity, ${\rm EW}_{\rm ion,\,fore}$ is treated as a constant fitting parameter for each ion, while the M\,31 contribution was calculated from the modified-$\beta$ density profile as a function of impact parameter.

To calculate ${\rm EW_{ion,\,M31}} (R_{\rm imp})$, we first integrated the model-predicted ion column density along the LOS, considering only gas within an assumed CGM boundary, $R_{\rm CGM}$:
\begin{equation}
N_{\rm ion}(R_{\rm imp}) =
\begin{cases}
\displaystyle \int_{-s_{\rm max}}^{s_{\rm max}} n_{\rm H}(r)\, f_{\rm ion}(T)\, f_{\rm O}\,(Z/Z_\odot)\, ds,
& R_{\rm imp} < R_{\rm CGM}, \\
0,
& R_{\rm imp} \ge R_{\rm CGM}.
\end{cases}
\end{equation}
Here, $n_{\rm H}(r)$ is the hydrogen number density of the hot halo of M\,31, described by the modified-$\beta$ model, $f_{\rm ion} (T)$ is the ion fraction of \osv or O\,{\sc viii}, $f_{\rm O}=5.37\times10^{-4}$ is the solar oxygen fraction relative to hydrogen \citep{Lodders2009}, and $Z=0.3~Z_\odot$ is the adopted metallicity. At our adopted temperature of $2.5\times10^6$~K, the \osv and \oeit fractions are $0.278$ and $0.448$ under CIE \citep{Gnat2007}. $s$ is the LOS coordinate measured from the point of closest approach to the M\,31 center, such that $s=0$ corresponds to $r=R_{\rm imp}$. The integration limits are set by the two intersections of the sightline with the adopted CGM boundary. The maximum LOS coordinate is:
\begin{equation}
s_{\rm max}=\sqrt{R_{\rm CGM}^2-R_{\rm imp}^2}.
\end{equation}
The transformation between galactocentric radius of M\,31 and $s$ is:
\begin{equation}
r(s)=\sqrt{R_{\rm imp}^2+s^2},
\end{equation}
 
The derived $N_{\rm ion}$ was then converted to the line EW using the curve-of-growth method. Specifically, we calculated the optical-depth profile, $\tau(\lambda)$, using the Voigt-profile model described by \citet{Buote2009}, assuming a Doppler-$b$ parameter of $120$~km\,s$^{-1}$. Throughout this paper, the adopted oscillator strengths and Einstein-$A$ coefficients for the Voigt profile are $0.696$ and $3.3\times10^{12}$~s$^{-1}$ for the \osv K$\alpha$ line, and $0.416$ and $2.57\times10^{12}$~s$^{-1}$ for the \oeit K$\alpha$, respectively \citep{Verner1996}. 
The EW can then be determined by integrating over the optical depth profile, with: 
\begin{equation}
{\rm EW} = \int [1 - e^{-\tau(\lambda)}] d\lambda.    
\end{equation}

\begin{figure}
\centering
\includegraphics[width=0.6\linewidth]{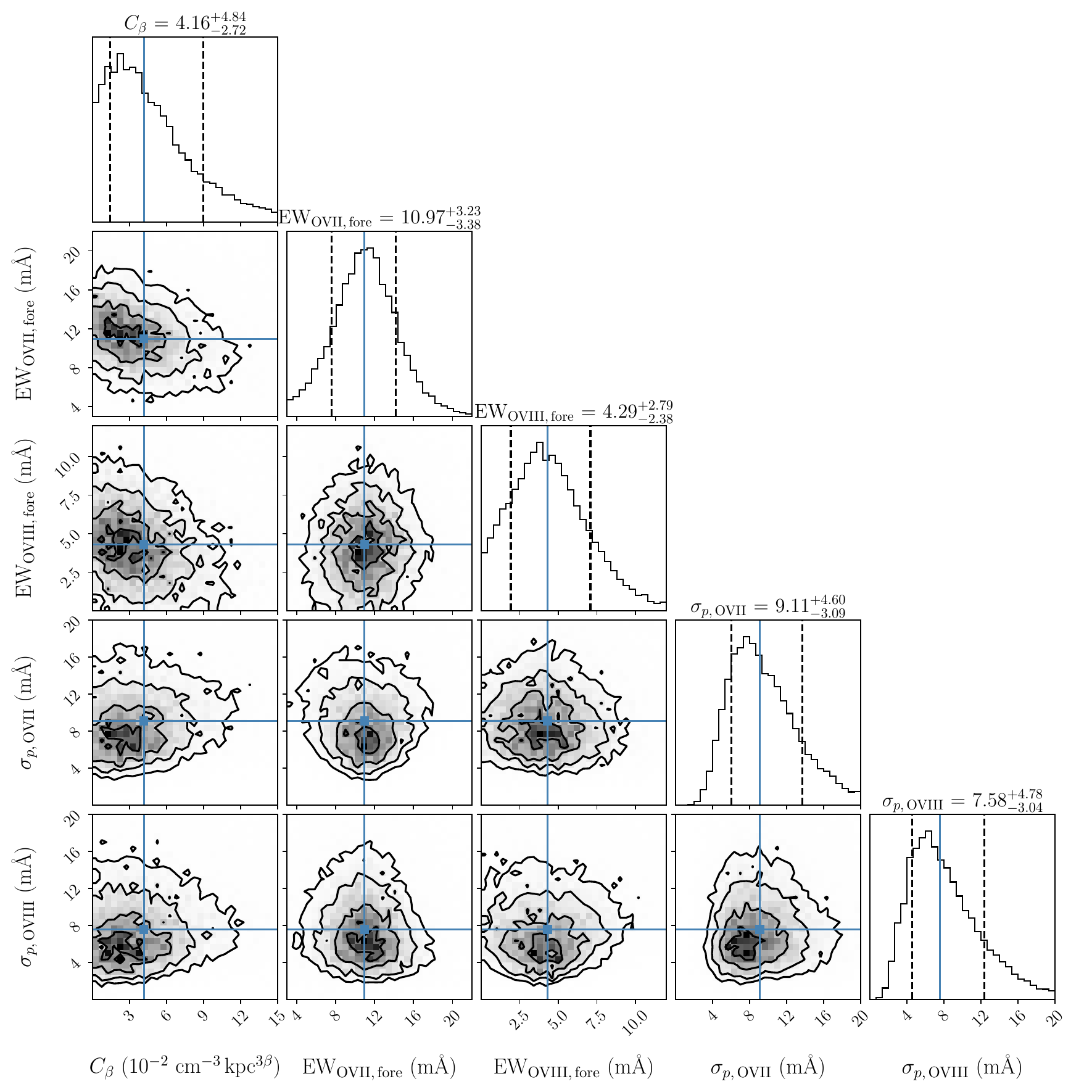}
\caption{Example posterior distributions of the gas-distribution model parameters, adopting a CGM boundary of $R_{\rm CGM}=400$~kpc.
}
\label{fig:examp_posterior}
\end{figure}

We fitted the model-predicted EWs to the observed EWs, using the Bayesian method described in Appendix~\ref{append:bayesian_method}. The \osv and \oeit data were fitted jointly, with separate foreground EWs and patchiness parameters for the two ions. We varied the assumed CGM boundary over $R_{\rm CGM}=300$--$500$~kpc in steps of $5$~kpc. Figure~\ref{fig:mhalo_rcgm} shows the inferred hot CGM mass of M\,31 as a function of the adopted CGM boundary. Figure~\ref{fig:examp_posterior} presents an example of the sampled posterior distribution for $R_{\rm CGM}=400$~kpc, plotted using the {\it Python} package {\it corner} \citep{Foreman-Mackey2016}.

\end{document}